\PassOptionsToPackage{table,svgnames,dvipsnames,rgb,hyperref}{xcolor}

\documentclass[]{style/ceurart}
\usepackage{listings}
\usepackage{graphicx}
\usepackage{tabularx}
\usepackage{placeins} 
\usepackage[justification=centering]{caption}
\usepackage{booktabs} 
\usepackage{amsmath}
\usepackage{subcaption}
\usepackage{tikz}
\usepackage{pgfplots}
\usetikzlibrary{bayesnet}

\begin{document}

\copyrightyear{2025}
\copyrightclause{Copyright for this paper by its authors.
  Use permitted under Creative Commons License Attribution 4.0
  International (CC BY 4.0).}

\conference{CLEF 2025 Working Notes, 9 -- 12 September 2025, Madrid, Spain}

\title{DS@GT at LongEval: Evaluating Temporal Performance in Web Search Systems and Topics with Two-Stage Retrieval}
\title[mode=sub]{Notebook for the LongEval Lab at CLEF 2025}

\author[1]{Anthony Miyaguchi}[
orcid=0000-0002-9165-8718,
email=acmiyaguchi@gatech.edu,
]
\cormark[1]
\author[2]{Imran Afrulbasha}[
orcid=0009-0002-7602-5194,
email=iafrulbasha3@gatech.edu,
]

\author[3]{Aleksandar Pramov}[
orcid=0009-0005-9049-1337,
email=apramov3@gatech.edu,
]

\address[1]{Georgia Institute of Technology, North Ave NW, Atlanta, GA 30332}
\cortext[1]{Corresponding author.}

\begin{abstract}
Information Retrieval (IR) models are often trained on static datasets, making them vulnerable to performance degradation as web content evolves. 
The DS@GT competition team participated in the Longitudinal Evaluation of Model Performance (LongEval) lab at CLEF 2025, which evaluates IR systems across temporally distributed web snapshots. 
Our analysis of the Qwant web dataset includes exploratory data analysis with topic modeling over time.
The two-phase retrieval system employs sparse keyword searches, utilizing query expansion and document reranking.
Our best system achieves an average NDCG@10 of 0.296 across the entire training and test dataset, with an overall best score of 0.395 on 2023-05.
The accompanying source code for this paper is at \url{https://github.com/dsgt-arc/longeval-2025}.
\end{abstract}


\maketitle

\section{Introduction}
Modern search engines operate on continually evolving corpora. 
However, most IR models are trained and validated on static datasets, leading to a phenomenon known as temporal drift, where effectiveness decreases as the test data diverges in time from the training data. 
Task 1 of CLEF's 2025 LongEval lab aims to address this issue by evaluating models across time-variant snapshots of web content. 
Compared to prior LongEval labs (2023 and 2024), this year's dataset is larger, with 18 million documents and 9,000 training queries.
The lag data is also more granular, providing monthly snapshots between June 2022 and August 2023.
The dataset is now French-only, in contrast to previous years, which provided English-translated documents.

Our team developed an IR pipeline to observe temporal variance using a two-phase retrieval system involving query expansion. 
We also leverage unsupervised topic modeling techniques to uncover latent thematic structures.
We submitted four distinct systems for evaluation, each representing a subset of the full feature set architecture:  

\begin{enumerate}
    \item \textbf{BM25 Baseline}: standard BM25 retrieval without any query expansion or reranking.
    \item \textbf{BM25 + Query Expansion}: standard BM25 retrieval with Gemini-based query expansion, but without reranking.
    \item \textbf{BM25 + Reranking}: standard BM25 retrieval with a Cross Encoder reranker, but without query expansion.
    \item \textbf{BM25 + Query Expansion + Reranking}: standard BM25 retrieval with Gemini query expansion, followed by a Cross Encoder reranker.
\end{enumerate}

\section{Related Work}

LongEval’s motivation is rooted in the observed decline in IR-model performance over time \cite{alkhalifa2024longeval}. 
An extensive survey documented this effect and concluded that increasing temporal distance degrades relevance, calling for retrieval models that incorporate temporal features \cite{campos2016survey}. 
Previous work connected retrieval accuracy to calendar cycles, showing that systems ignoring weekly and yearly periodicities systematically under-rank timely documents—such as sports fixtures or fiscal reports \cite{keikha2011time}.

Previous LongEval research quantified the degradation and proposed mitigations, including frequent model updates and query-time reranking \cite{alkhalifa2024longeval}. 
The present work extends that line by first characterizing the upstream data through topic modeling—a method explored in web-search retrieval \cite{wei2006lda,ai2016contextual}—and then presenting a more resilient architecture for large-scale experimentation.

The approach adopted in our work follows the standard two-stage pipeline in which a sparse BM25 retriever is followed by a neural re-ranker \cite{hambarde2023information,clavié2024rerankers}.
Because automatic query expansion has repeatedly improved retrieval effectiveness \cite{carpineto2012survey}, the system queries Google’s Gemini LLM \cite{GoogleGemini} to generate expansions before retrieval.

\section{Qwant Search Engine Dataset}

\subsection{Shared Task Data Collection}

We briefly touch upon the acquisition of pages and queries from the commercial search engine Qwant as part of the shared task setup. 
The approach largely follows the Cranfield Paradigm, where data acquisition is conducted periodically, forming a sequence of sub-collections \cite{galuscakova2023longeval}. 
The data collection process involves constructing topics, queries, relevance estimates, and documents.

Topics are selected once and shared across all subsequences guided by several criteria \cite{galuscakova2023longeval}. 
First, topics needed to be popular enough to generate a substantial number of relevant queries. 
Second, they had to be stable over time to support longitudinal performance evaluation of the information retrieval (IR) system. 
The persistence of these topics across different periods is assessed with tools like Google Trends \cite{galuscakova2023longeval}. 
Lastly, the topics were required to be general enough to encompass a wide variety of queries. 
The final set of topics chosen by the dataset organizers is shown in Table~\ref{tab:topics}.

The query selection process begins with extracting user topics and mapping them to real queries answered by Qwant's search engine, ensuring all displayed results are indexed. 
Queries are matched to topics using substring filtering, forming sets of relevant queries per topic. 
Since this process can generate tens of thousands of queries, a top-k selection retains only the most frequently asked queries per topic, reducing redundancy \cite{galuscakova2023longeval}. 

\begin{equation}
Q_{\mathcal{T}} = \bigcup_{t \in \mathcal{T}} Q_t 
    \quad\text{such that}\quad 
Q_t = \{ q \mid q \in Q, t \subseteq_{\text{str}} q \}
\end{equation}

If \text{Q} is the set of all Qwant queries and \text{T} is the set of topics defined in Table 1, for each topic \( t \in \mathcal{T} \) the lab organizers select all the queries \( Q_t \) from \text{Q} that contain \text{t} as a sub-string \cite{galuscakova2023longeval}. 


Next, queries undergo automatic filtering, selecting only those with at least 10 relevance assessments, followed by a manual review to merge similar queries and remove adult content \cite{galuscakova2023longeval}. 
This structured filtering refines the dataset, ensuring more reliable query distributions.

\begin{table}[h]
    \centering
    \renewcommand{\arraystretch}{0.4} 
    \setlength{\tabcolsep}{2pt} 
    \begin{tabular}{|c| l| l| c| l| l|}
        \hline
        \scriptsize
        \textbf{No.} & \textbf{Topic} & \textbf{English Description} & \textbf{No.} & \textbf{Topic} & \textbf{English Description} \\
        \hline
        1  & Eau & Water & 15 & Président & President \\
        2  & Nourriture & Food & 16 & Pétrole & Oil \\
        3  & Espace & Space & 17 & Impôts & Taxes \\
        4  & Voiture & Car & 18 & Votants & Voters \\
        5  & Argent & Money & 19 & Fraude & Fraud \\
        6  & Manifestation & Protest & 20 & Élisabeth Borne & (French PM, May 2022) \\
        7  & Virus & Virus & 21 & Changement climatique & Climate change \\
        8  & Terre & Earth & 22 & Fête du Travail & (French holiday) \\
        9  & Énergie & Energy & 23 & Eurovision & Eurovision \\
        10 & Police & Police & 24 & Jacques Perrin & (French actor) \\
        11 & Loi & Law & 25 & Régine & (French singer) \\
        12 & Travailleurs & Workers & 26 & Heartstopper & - \\
        13 & Guerre & War & 27 & Fête des Mères & Mother's Day \\
        14 & Invasion & Invasion & 28 & Johnny Depp & - \\
        \hline
    \end{tabular}
    \caption{French topics and their English descriptions}
    \label{tab:topics}
\end{table}
\FloatBarrier 

The relevance estimates for LongEval-Retrieval rely on implicit feedback from user clicks, as Qwant preserves privacy by not tracking multiple clicks, dwell times, or query reformulations \cite{galuscakova2023longeval}. 
Since raw click data is noisy and biased toward top-ranked results, Click Models are used to infer document relevance while minimizing bias. 
Given Qwant's privacy constraints, a Cascade Model—a simplified version of Dynamic Bayesian Networks (DBN)—is employed, where users scan results from top to bottom and click only on attractive documents \cite{galuscakova2023longeval}. 
The attractiveness parameter (\(\alpha\)) is estimated through Maximum Likelihood Estimation (MLE), providing a probabilistic relevance measure for query-document pairs. 
To make these relevance estimates compatible with traditional IR metrics, \(\alpha\) is mapped to discrete relevance values (0 = not relevant, 1 = relevant, 2 = highly relevant). 
The MLE is given by the following form for a query \text{q} and document \text{d} where \( S_{q,d} \) denotes the set of all instances in Qwant's query log in which document \( d \) was displayed in a search engine results page (SERP) at or above the rank of a clicked document. 
The binary variable \( c_s(d) \) indicates whether document \( d \) was clicked within a given entry \(S\).

\begin{equation}
\hat{\alpha}_{q,d} = \frac{1}{|S_{q,d}|} \sum_{s \in S_{q,d}} c_s(d)
\end{equation}

The document corpus for LongEval-Retrieval is extracted from Qwant's search index, including both the documents displayed in search results and a random sample of non-relevant documents to minimize bias. 
To avoid skewing the corpus toward Qwant's ranking function, up to 100,000 documents per topic are randomly selected based on matching word tokens. 
Before inclusion, documents undergo a cleaning process, in which their text is extracted using Qwant's internal tools and filtered for adult and spam content.

\subsection{Train Dataset Statistics}
\label{sec:dataset-stats}

We provide statistics over the training dataset to aid decisions related to cost-effectiveness and evaluation time.
We use the tiktoken library to tokenize documents alongside a naive whitespace tokenizer.
In table~\ref{tab:longeval_avg_per_doc}, we see that there about two million documents per time-step, with an average of 850 words or 1,300 tokens per document.
We provide per-million (PM) count statistics which can be used in throughput to cost or time calculations in table~\ref{tab:longeval_sum_by_date_sorted}. and table~\ref{tab:longeval_sum_all}.

This data is useful to extrapolating dense-retrieval workflows.
Full-scale embedding using self-hosted encoder-only sentence transformers require almost an entire day of GPU hours.
With the \texttt{nomic-ai/nomic-embed-text-v2-moe} encoder on the train dataset for 2022-08, we run 100\% utilization against a v100 GPU in 50 parts at 25 minutes a part for a total of 21 hours of GPU time.

We can use these statistics to estimate the cost of what it might take to process all the data with a large language model (LLM).
For example, as of May 2025, the Google Gemini 2.0 model costs 10 cents per million tokens of input and 40 cents per million tokens of output.
The projected cost of reading the entire training dataset as input context would be \$2,528 USD, which is cost-prohibitive for experimentation.

\begin{table}[htbp]
  \centering
  \caption{LongEval Dataset Statistics: Overall Statistics for the French Training Split.}
  \label{tab:longeval_avg_per_doc}
  \begin{tabular}{@{}llrrrrrr@{}}
    \toprule
    Split & Language & Count & Sum Tokens & Avg Tokens & Stddev Tokens & Avg Words & Stddev Words \\
    \midrule
    train & French & 19000580 & $2.53 \times 10^{10}$ & 1330.3 & 889.1 & 849.7 & 549.7 \\
    \bottomrule
  \end{tabular}
\end{table}

\begin{table}[htbp]
  \centering
  \caption{LongEval Dataset Statistics: Sum over each date (French, train split).}
  \label{tab:longeval_sum_by_date_sorted}
  \begin{tabular}{@{}lllrrrr@{}}
    \toprule
    Split & Language & Date    & Sum Tokens          & Sum Words           & Sum Tokens PM & Sum Words PM \\
    \midrule
    train & French   & 2022-06 & $2.38 \times 10^9$  & $1.51 \times 10^9$  & 2378.2        & 1514.5       \\
    train & French   & 2022-07 & $2.38 \times 10^9$  & $1.52 \times 10^9$  & 2379.1        & 1515.3       \\
    train & French   & 2022-08 & $2.40 \times 10^9$  & $1.53 \times 10^9$  & 2396.1        & 1526.0       \\
    train & French   & 2022-09 & $1.65 \times 10^9$  & $1.05 \times 10^9$  & 1650.2        & 1052.6       \\
    train & French   & 2022-10 & $3.35 \times 10^9$  & $2.14 \times 10^9$  & 3349.5        & 2136.6       \\
    train & French   & 2022-11 & $3.37 \times 10^9$  & $2.15 \times 10^9$  & 3369.2        & 2148.7       \\
    train & French   & 2022-12 & $3.26 \times 10^9$  & $2.09 \times 10^9$  & 3262.8        & 2091.1       \\
    train & French   & 2023-01 & $3.27 \times 10^9$  & $2.09 \times 10^9$  & 3268.7        & 2094.8       \\
    train & French   & 2023-02 & $3.22 \times 10^9$  & $2.07 \times 10^9$  & 3222.3        & 2065.1       \\
    \bottomrule
  \end{tabular}
\end{table}

\begin{table}[htbp]
  \centering
  \caption{LongEval Dataset Statistics: Sum over all collections (French, train split).}
  \label{tab:longeval_sum_all}
  \begin{tabular}{@{}llrrrr@{}}
    \toprule
    Split & Language & Sum Tokens            & Sum Words             & Sum Tokens PM & Sum Words PM \\
    \midrule
     train & French   & $2.53 \times 10^{10}$ & $1.61 \times 10^{10}$ & 25276.2       & 16144.7      \\
    \bottomrule
  \end{tabular}
\end{table}

\section{Topic Modeling for Exploratory Data Analysis}

Topic modeling is a machine learning technique used to uncover latent themes or patterns from large datasets, providing insight into the evolving structure of textual information. 
In the context of LongEval, we leveraged topic modeling to analyze how thematic distributions shift over time, capturing longitudinal trends in digital content. 
Recognizing non-trivial topic drifts could provide some insight into any search engine performance fluctuations.

\subsection{Topic Models}

\subsubsection{Non-negative Matrix Factorization }

Non-Negative Matrix Factorization (NMF) factorizes a given non-negative matrix \( \mathbf{X} \) into two lower-dimensional non-negative matrices, \( \mathbf{W} \) and \( \mathbf{H} \) \cite{kassab2024fairerNMF}. 
This factorization is particularly useful for topic modeling, as it provides an interpretable structure where documents are mixtures of topics characterized by distributions over words. 
By factorizing the document-term matrix, we can capture the underlying themes present in long-term information retrieval datasets, helping to evaluate how topic distributions shift over time. 



The optimization objective for NMF seeks to minimize the reconstruction error between \( \mathbf{X} \) and \( \mathbf{W} \times \mathbf{H} \) \cite{scikit-learn-nmf}. 
\( \mathbf{X} \in \mathbb{R}^{m \times n} \) represents the document-term matrix, with \( m \) denoting the total number of documents in the corpus and \( n \) representing the number of unique terms in the vocabulary. 
The matrix \( \mathbf{W} \in \mathbb{R}^{m \times k} \) captures the document-topic distribution, where each row indicates the strength of topic presence in a document. 
Meanwhile, \( \mathbf{H} \in \mathbb{R}^{k \times n} \) defines the topic-word associations, with each row highlighting the key terms that define a given topic. 

The NMF objective is as follows:

\[
\min_{\mathbf{W} \geq 0, \mathbf{H} \geq 0} \quad f(\mathbf{W}, \mathbf{H}) = \|\mathbf{X} - \mathbf{W} \mathbf{H}\|^2_F
\]

where \( \|\cdot\|^2_F \) denotes the Frobenius norm.
The constraints \( \mathbf{W} \geq 0 \) and \( \mathbf{H} \geq 0 \) enforce non-negativity, ensuring that topics and their respective word distributions remain interpretable \cite{park_nmf}. 
The Frobenius norm ensures smooth optimization, enabling faster convergence with gradient-based methods. 
Unlike the generalized Kullback-Leibler (KL) divergence, it does not require a probabilistic interpretation of the input matrix, making it more suitable for general-purpose topic modeling. 

\subsubsection{Latent Dirichlet Allocation }

LDA is a probabilistic framework for unsupervised topic discovery \cite{wei2006lda}. 
It assumes that each document is composed of multiple latent topics, with each topic represented by a distribution over words. 
The following joint probability distribution governs the complete generative process in LDA.

\[
p(\beta_{1:K}, \theta_{1:D}, z_{1:D}, w_{1:D}) = \prod_{i=1}^{K} p(\beta_i) \prod_{d=1}^{D} p(\theta_d) \left( \prod_{n=1}^{N} p(z_{d,n} | \theta_d) p(w_{d,n} | \beta_{1:K}, z_{d,n}) \right)
\]

where:
\begin{itemize}
  \item \( \beta_{1:K} \) are the topic distributions over the vocabulary.
  \item \( \theta_{1:D} \) are the document-specific topic proportions.
  \item \( z_{d,n} \) is the topic assignment for the \(n\)the word in document \(d\), drawn from \( \text{Multinomial}(\theta_d) \).
  \item \( w_{d,n} \) is the observed word, drawn from the topic \( z_{d,n} \)'s word distribution \( \beta_{z_{d,n}} \).
\end{itemize}

This joint distribution reflects the conditional dependencies among the variables:
\begin{itemize}
  \item The topic assignment \( z_{d,n} \) depends on the document-specific topic proportions \( \theta_d \).
  \item The observed word \( w_{d,n} \) depends on the topic assignment \( z_{d,n} \) and all the topics \( \beta_{1:K} \).
\end{itemize}

\subsection{Exploratory Data Analysis}

\subsubsection{Methodology}

Our modeling pipelines for NMF and LDA both leverage Luigi\cite{luigi} for orchestration, PySpark (SparkSQL/MLlib) for preprocessing, and scikit-learn for end-to-end model execution \cite{scikit-learn-nmf}. We randomly sample documents from the entire Qwant collection to ensure a diverse subset for topic modeling while reducing computational load.
The text preprocessing stage involves tokenization followed by conversion into a document-term matrix using a term frequency representation.
Next, both NMF and LDA topic models are trained with 20 topics to extract topic distributions across documents. 
An LLM summarizes the 100 top words for each topic, specifically Grok 3, a 2.7 trillion parameters LLM.

To visualize NMF results, the topic distributions of a random subset of documents within an arbitrary month were inferred using a pre-trained NMF model, yielding document-topic association scores. 
For both NMF and LDA, we projected the high-dimensional topic embeddings into two dimensions to facilitate clustering analysis using principal component analysis (PCA) and Gaussian random projections (GRP).
We create a scatter plot of the projections, with colors indicating the dominant topic assignments per document. 
These visualizations offer insight into the topic structure uncovered by both modeling methods while preserving essential document relationships in a reduced space.

\subsubsection{Discussion}

\begin{figure}[htbp]
    \centering
    \label{fig:topic_distributions_combined}

    \begin{subfigure}{\columnwidth}
        \centering
        \caption{Topic Groupings from June 2022 to October 2022}
        \label{tab:topic_distributions_summer}
        \begin{tabularx}{\linewidth}{|X|X|X|}
            \hline
            \multicolumn{3}{|c|}{\textbf{Non-negative Matrix Factorization Topic Groupings}} \\ 
            \hline
            \textbf{2022\_06} & \textbf{2022\_08} & \textbf{2022\_10}  \\ 
            \hline
            French General History  & French General History & French Urban Development  \\
            English News and Politics  & English General History  & English News and Discourse  \\
            French Media production  & French Media  & French Social Media  \\
            Online Forums & French Social Media & Online Pharmacies \\
            Online Pharmacies & Online Pharmacies & French E-commerce \\
            French E-commerce & E-commerce & Archival Content \\
            Blog Archives & Blog Archives & Film Industry and Awards \\
            French politics & European Elections & Hotel Bookings and Travel \\
            French Public Administration & French Administration & French Job Market \\
            English Online Engagement & English Social Media & English Social Media \\
            Global Development  & Global Development & Celebrity News \\
            French Elections  & French Elections & French Politics \\
            Website Analytics  & Regional News & French Fuel Prices \\
            French News & Book Subscriptions & Countries and Financial Institutions \\
            Book Resources & Website Privacy & Novelty Badges and Gifts \\
            Sustainable development & Sustainability & Restaurants and Cuisines \\
            Travel Accommodations & Travel &  Professional Names and Consulting \\
            Auctions & Auctions & French Administrative Processes \\
            French Local Governance & Local Governance & Product Listings and Transactions \\
            Social Impact & Entertainment & Book Pricing and Newsletters \\
            \hline
        \end{tabularx}
    \end{subfigure}

    \vspace{2em} 

    \begin{subfigure}{\columnwidth}
        \centering
        \caption{Topic Groupings for December 2022 and February 2023}
        \label{tab:topic_distributions_winter}
        \begin{tabularx}{\linewidth}{|X|X|}
            \hline
            \multicolumn{2}{|c|}{\textbf{Non-negative Matrix Factorization Topic Groupings}} \\ 
            \hline
            \textbf{2022\_12} & \textbf{2023\_02} \\ 
            \hline
            French Public Services and Environment & Workplace Policies and Enterprise management \\
            English News and History & Historical Events and Politics \\
            Online Pharmacies & Online pharmacies and Health \\
            French E-commerce & Home Improvement and Vehicle Maintenance \\
            Countries and Geopolitical Regions & Countries and Geopolitical Regions \\
            French Personal Communication & Personal Narratives and Societal Issues \\
            Hotel Bookings and Pricing & Hotel bookings and Travel Accommodations  \\
            English Social Media & Social Media Interaction \\
            Events and Eurovision content & Financial Institutions and International Entities  \\
            International Organizations, and Finance & Events and Historical Records \\
            Film Awards and Festivals & Job Opportunities and Recruitment \\
            French Jobs and Recruitment & Film Awards and Festivals \\
            Data privacy and Account Management & User Account Management, Data Privacy \\
            French Politics and Global Affairs & E-commerce and Customer Service \\
            Vehicle Sales and Real Estate & Vehicle Specifications and Real Estate \\
            French Fuel Prices and Regional Departments & Fuel prices and French Departments \\
            French Administrative Processes and Health & Administrative Processes and Legal Requirements \\
            Website cookies and User Analytics & Travel Bookings and Product Sales \\
            Product pricing and Shipping &  Online Forums and Personal Discussions \\
            Book Pricing and Newsletter Subscriptions & Energy Transition and Corporate Sustainability \\
            \hline
        \end{tabularx}
    \end{subfigure}
    \caption{20 Topic Groupings using Non-negative Matrix Factorization}
\end{figure}

\begin{figure}[htbp]
    \centering
    \begin{subfigure}{\columnwidth}
        \centering
        \resizebox{\linewidth}{!}{%
            \begin{subfigure}{0.19\textwidth}
                \includegraphics[width=\linewidth, trim={60 30 180 50}, clip]{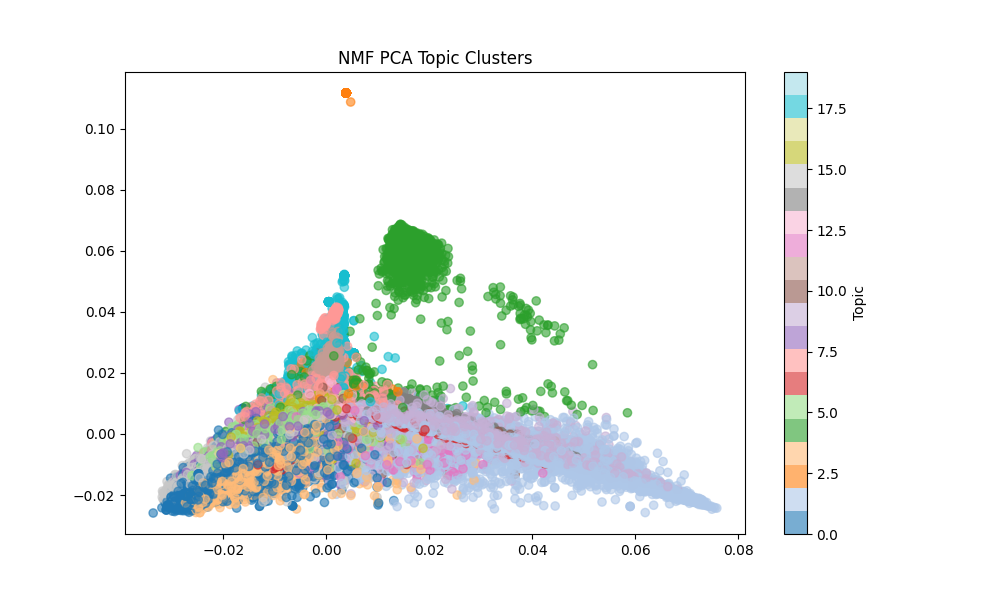}
            \end{subfigure}
            \begin{subfigure}{0.19\textwidth}
                \includegraphics[width=\linewidth, trim={60 30 180 50}, clip]{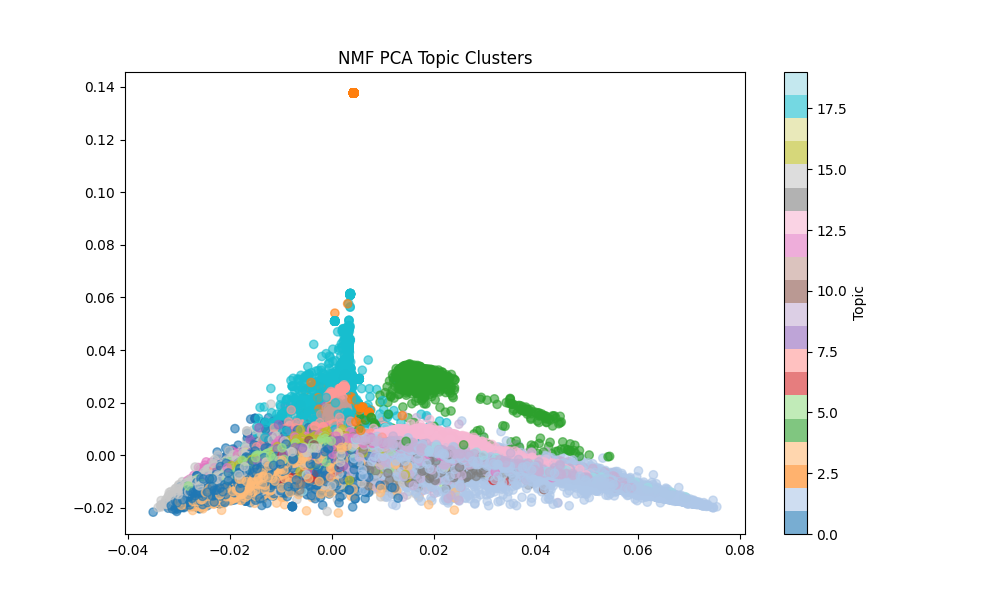}
            \end{subfigure}
            \begin{subfigure}{0.19\textwidth}
                \includegraphics[width=\linewidth, trim={60 30 180 50}, clip]{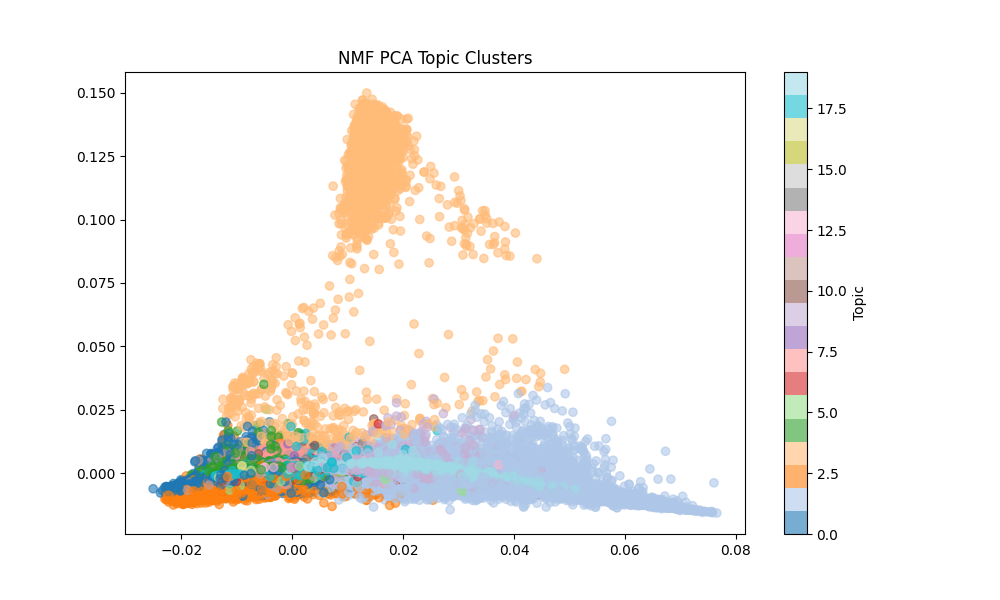}
            \end{subfigure}
            \begin{subfigure}{0.19\textwidth}
                \includegraphics[width=\linewidth, trim={60 30 180 50}, clip]{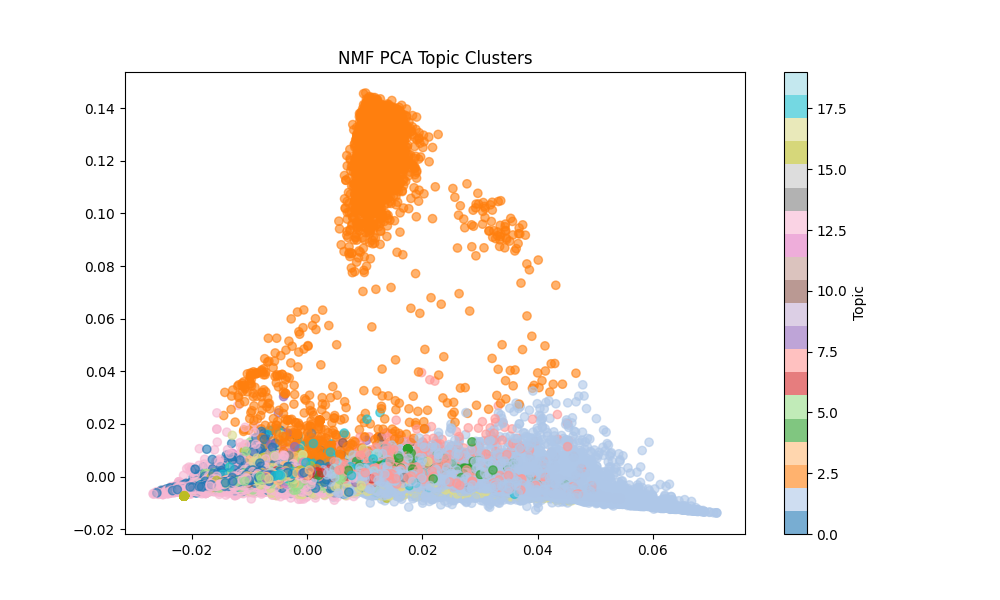}
            \end{subfigure}
            \begin{subfigure}{0.19\textwidth}
                \includegraphics[width=\linewidth, trim={60 30 180 50}, clip]{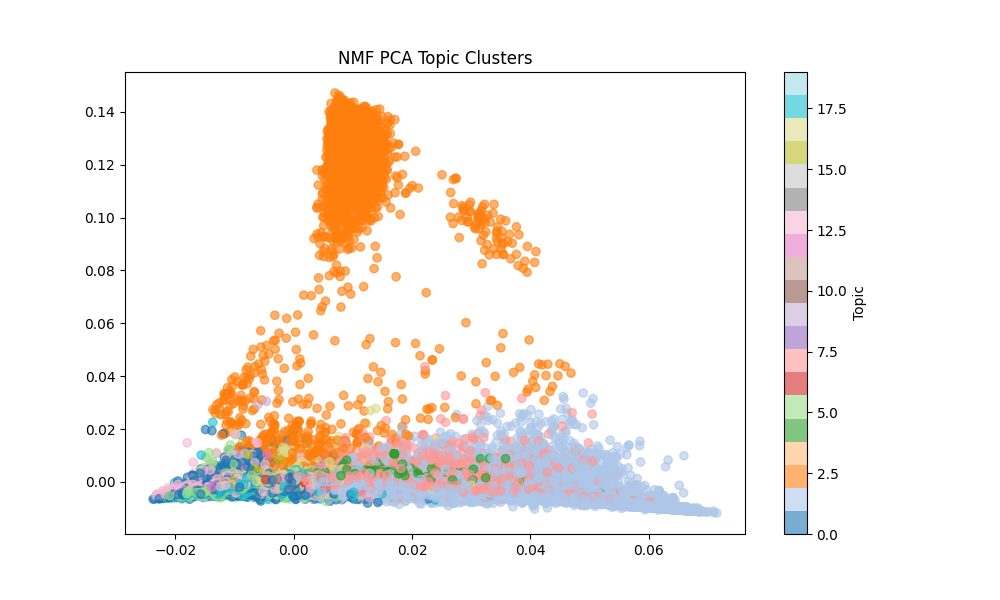}
            \end{subfigure}
        }
        \caption{NMF PCA topic distributions.}
        \label{fig:nmf_pca_row}
    \end{subfigure}
    
    \vspace{1em} 

    \begin{subfigure}{\columnwidth}
        \centering
        \resizebox{\linewidth}{!}{%
            \begin{subfigure}{0.19\textwidth}
                \includegraphics[width=\linewidth, trim={60 30 180 50}, clip]{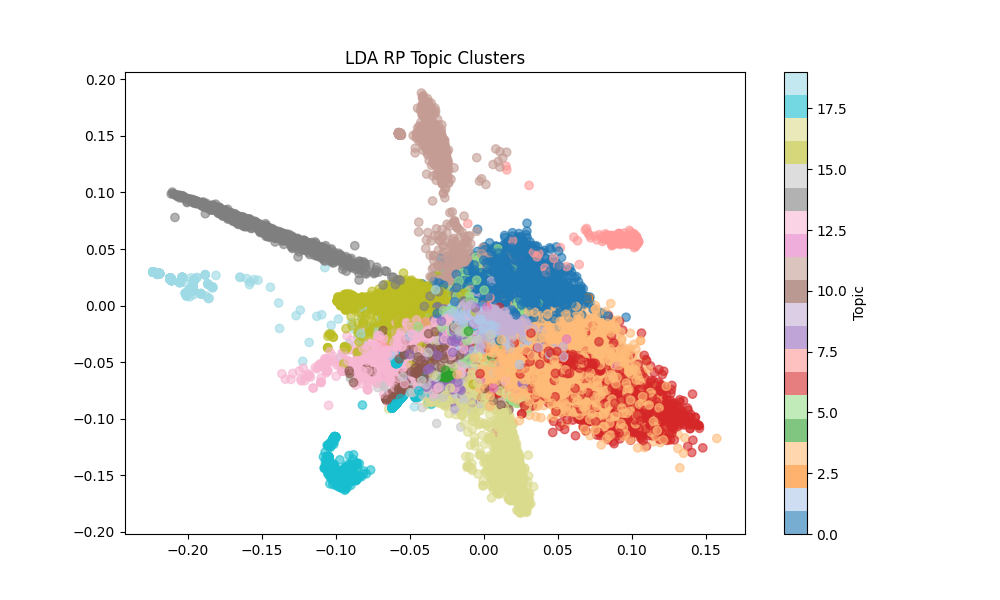}
            \end{subfigure}
            \begin{subfigure}{0.19\textwidth}
                \includegraphics[width=\linewidth, trim={60 30 180 50}, clip]{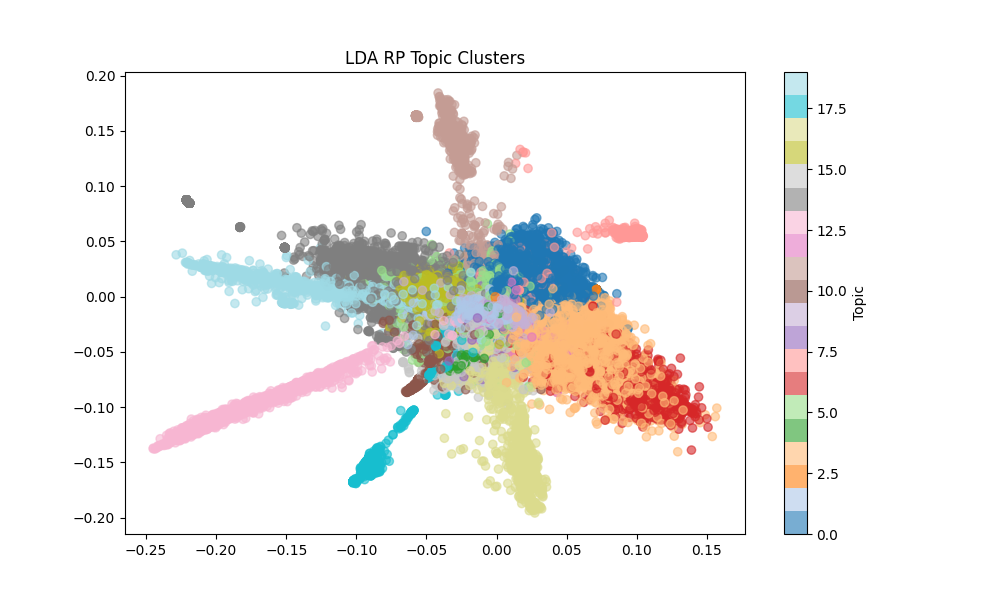}
            \end{subfigure}
            \begin{subfigure}{0.19\textwidth}
                \includegraphics[width=\linewidth, trim={60 30 180 50}, clip]{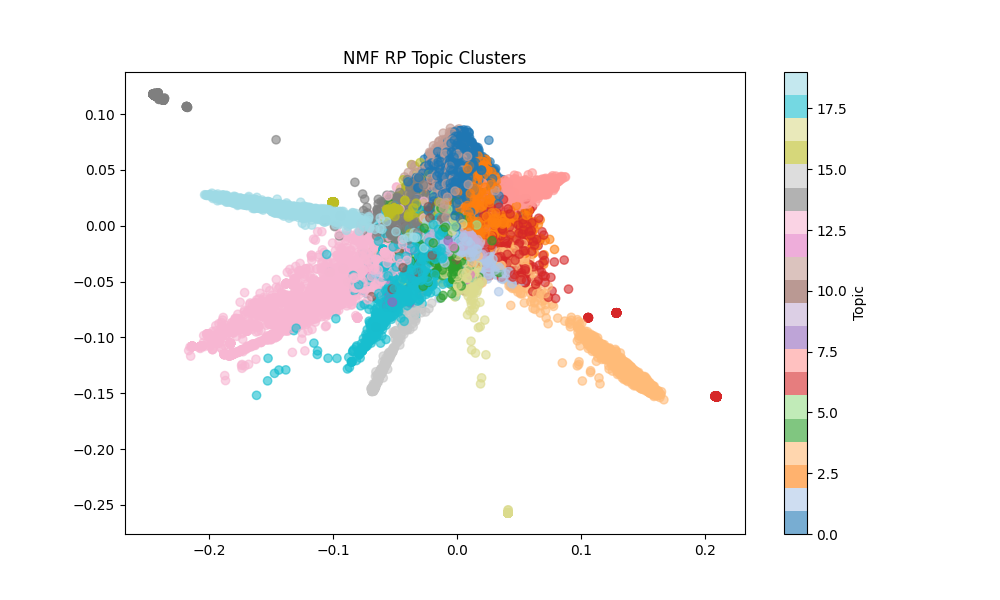}
            \end{subfigure}
            \begin{subfigure}{0.19\textwidth}
                \includegraphics[width=\linewidth, trim={60 30 180 50}, clip]{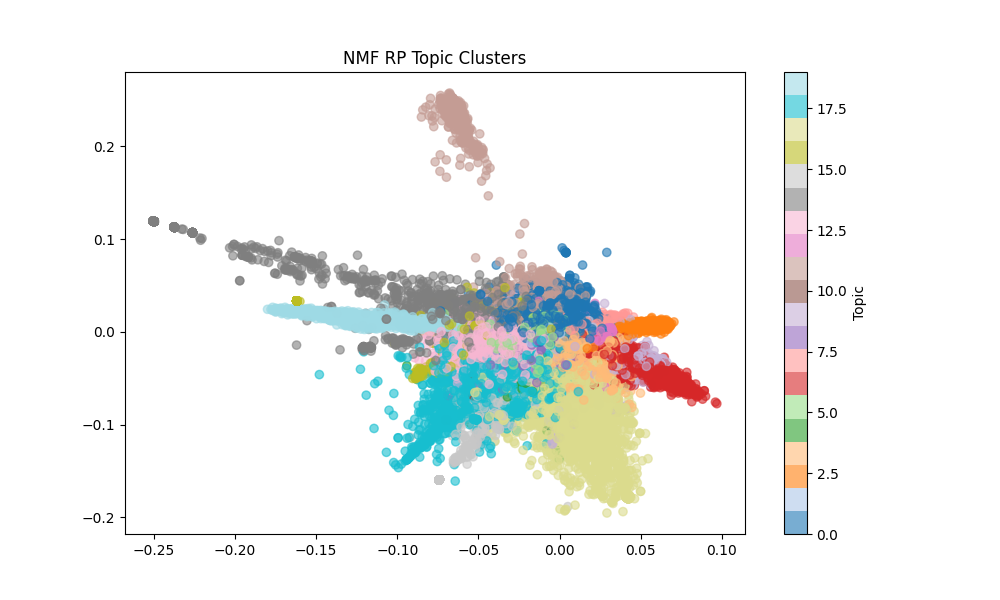}
            \end{subfigure}
            \begin{subfigure}{0.19\textwidth}
                \includegraphics[width=\linewidth, trim={60 30 180 50}, clip]{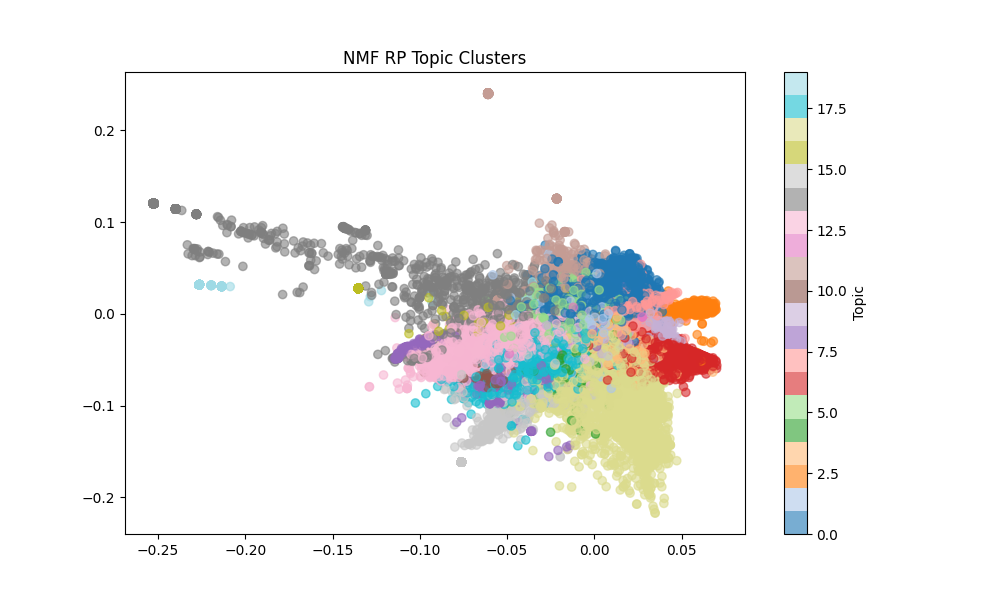}
            \end{subfigure}
        }
        \caption{NMF GRP topic distributions.}
        \label{fig:nmf_grp_row}
    \end{subfigure}
    
    \vspace{1em} 

    \begin{subfigure}{\columnwidth}
        \centering
        \resizebox{\linewidth}{!}{%
            \begin{subfigure}{0.19\textwidth}
                \includegraphics[width=\linewidth, trim={60 30 180 50}, clip]{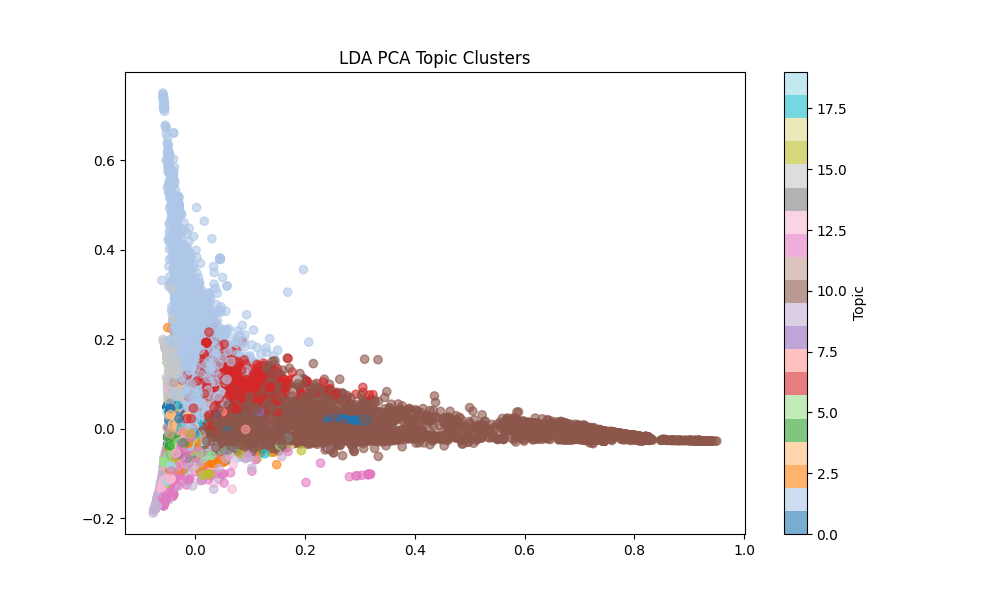}
            \end{subfigure}
            \begin{subfigure}{0.19\textwidth}
                \includegraphics[width=\linewidth, trim={60 30 180 50}, clip]{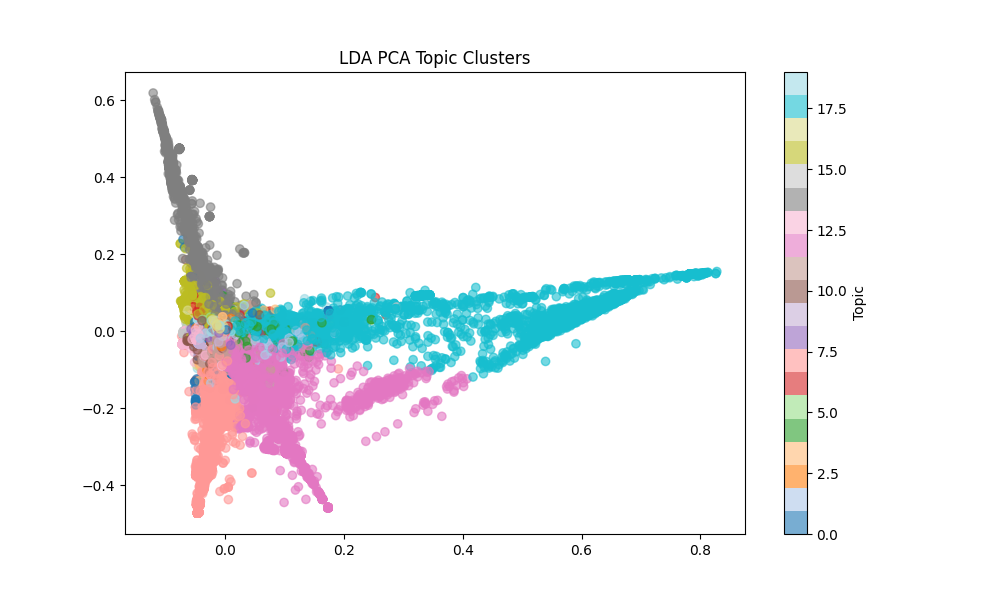}
            \end{subfigure}
            \begin{subfigure}{0.19\textwidth}
                \includegraphics[width=\linewidth, trim={60 30 180 50}, clip]{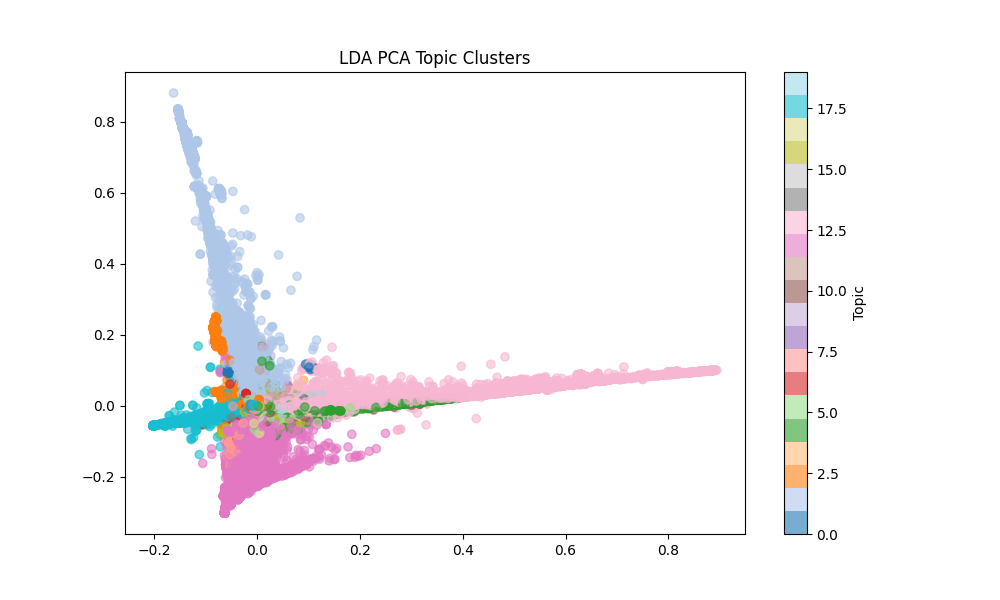}
            \end{subfigure}
            \begin{subfigure}{0.19\textwidth}
                \includegraphics[width=\linewidth, trim={60 30 180 50}, clip]{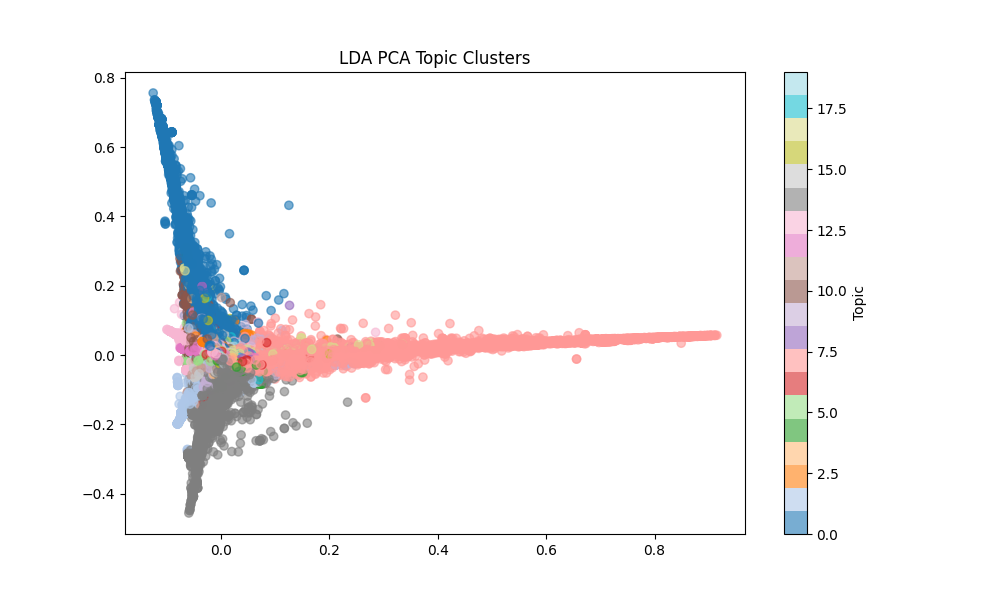}
            \end{subfigure}
            \begin{subfigure}{0.19\textwidth}
                \includegraphics[width=\linewidth, trim={60 30 180 50}, clip]{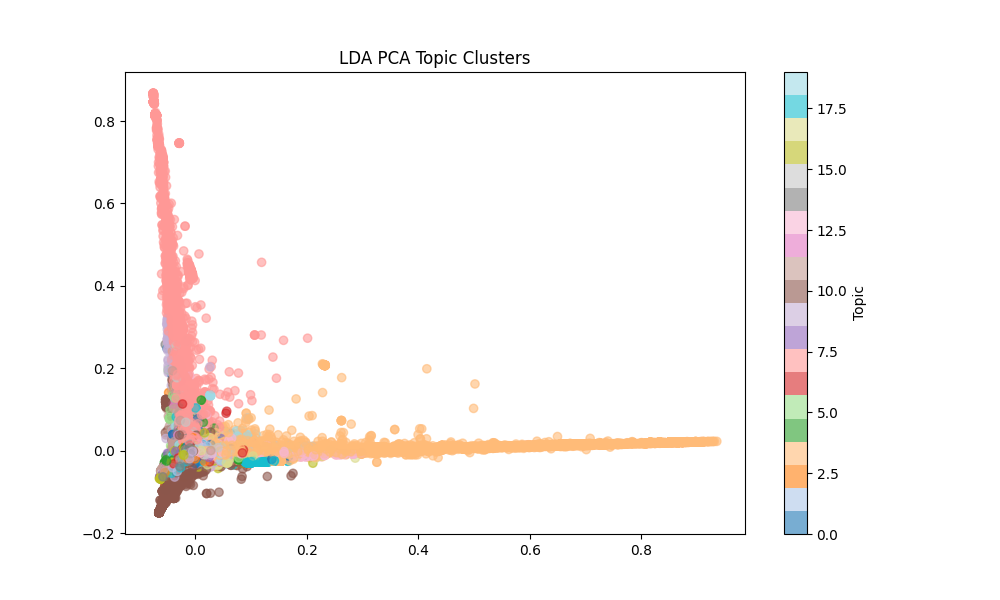}
            \end{subfigure}
        }
        \caption{LDA PCA topic distributions.}
        \label{fig:lda_pca_row}
    \end{subfigure}
    
    \vspace{1em} 

    \begin{subfigure}{\columnwidth}
        \centering
        \resizebox{\linewidth}{!}{%
            \begin{subfigure}{0.19\textwidth}
                \includegraphics[width=\linewidth, trim={60 30 180 50}, clip]{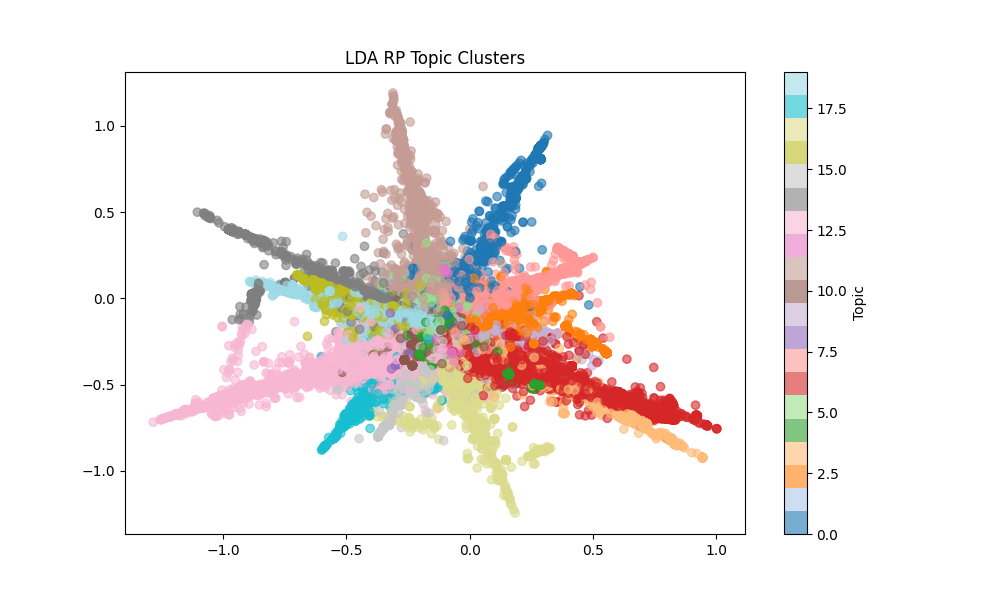}
            \end{subfigure}
            \begin{subfigure}{0.19\textwidth}
                \includegraphics[width=\linewidth, trim={60 30 180 50}, clip]{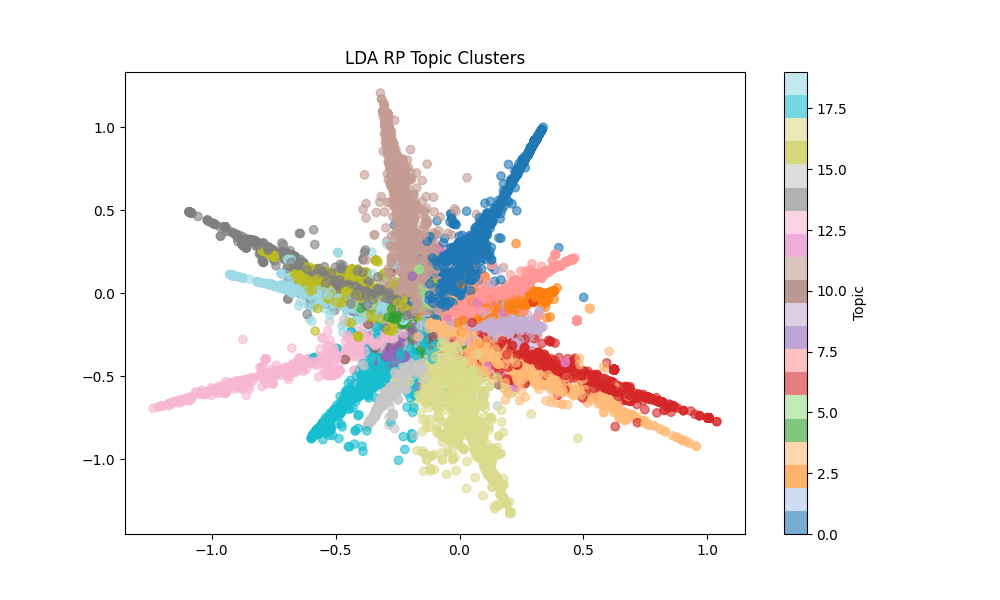}
            \end{subfigure}
            \begin{subfigure}{0.19\textwidth}
                \includegraphics[width=\linewidth, trim={60 30 180 50}, clip]{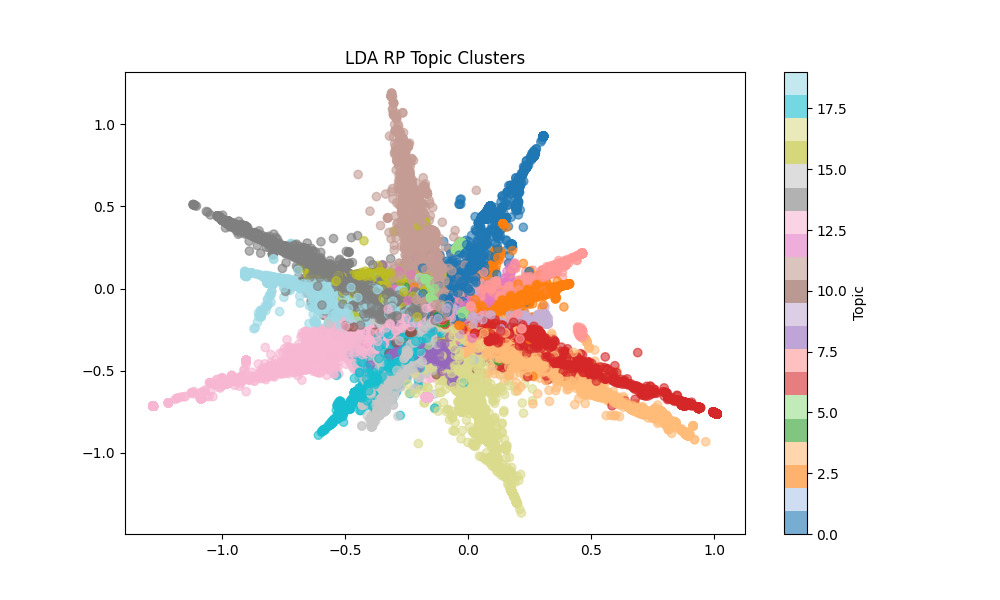}
            \end{subfigure}
            \begin{subfigure}{0.19\textwidth}
                \includegraphics[width=\linewidth, trim={60 30 180 50}, clip]{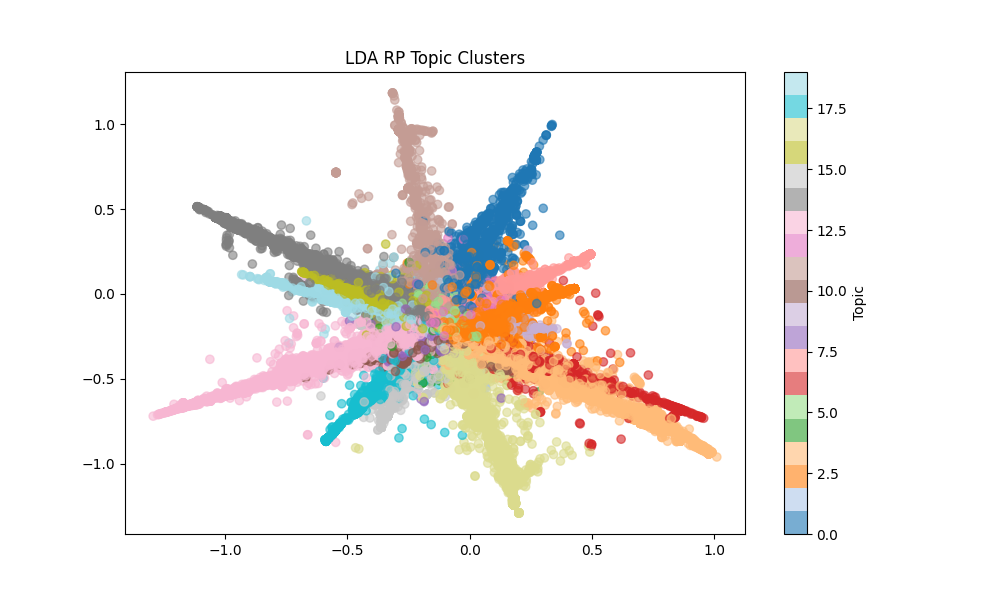}
            \end{subfigure}
            \begin{subfigure}{0.19\textwidth}
                \includegraphics[width=\linewidth, trim={60 30 180 50}, clip]{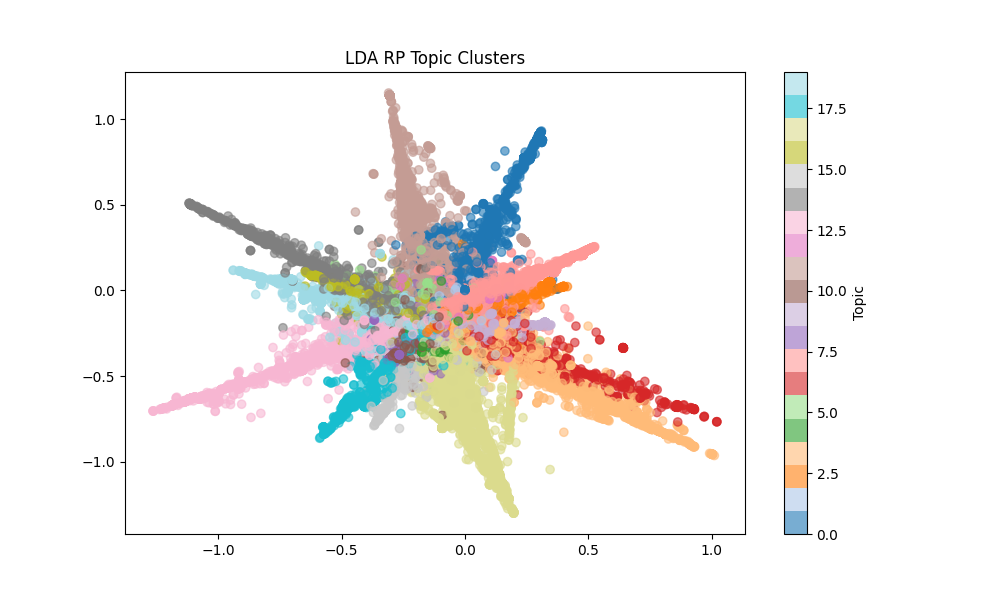}
            \end{subfigure}
        }
        \caption{LDA GRP topic distributions.}
        \label{fig:lda_grp_row}
    \end{subfigure}

    \caption{
    Comparison of topic model distributions from June 2022 to February 2023 in two-month intervals, i.e., 2022-06, 2022-08, 2022-10, 2022-12, and 2023-02.
    Each subfigure (a-d) shows the temporal evolution of topics for a given model (NMF/LDA) and dimensionality reduction technique (PCA/GRP).
    }
    \label{fig:combined_stacked}
\end{figure}

Figures \ref{tab:topic_distributions_summer} and \ref{tab:topic_distributions_winter} show topic groupings for equally spread out subsequences in the overall collection spanning June 2022 to February 2023. 
These topics are coherent and cover subjects one might expect to search for, including public services, e-commerce, social media, jobs, real estate, and travel.
The topics also capture temporal saliency through current events, such as fluctuations in fuel prices and upcoming elections.

\begin{table}[h]
    \centering
    \caption{Structural Fillers During LDA Runs}
    \begin{tabular}{|c|l|}
        \hline
        \multicolumn{2}{|c|}{\textbf{Structural Fillers Surfaced From LDA}} \\ 
        \hline
        \textbf{Category} & \textbf{Examples} \\
        \hline
 Common Articles & de, la, le, les, une, un, des, du \\
 Prepositions & à, en, dans, sur, sous, entre, pour, avec, sans \\
 Pronouns & je, vous, il, elle, nous, vous, ils, mon, son, leur \\
 Conjunctions & et, ou, mais, si, que, donc, car, ni, or \\
 Verbs (Common Auxiliary) & être, peut, a, ont, fait, c’est, était \\
 Punctuation & ., ,, ?, !, -, :, /, |, €, «, » \\
        \hline
    \end{tabular}
    \label{tab:filler}
\end{table}

In terms of French text preprocessing, our LDA topic summary documents display a substantial amount of filler text, as shown in Table~\ref{tab:filler}. A subsequent attempt at document cleanup might involve using NLTK's WordNetLemmatizer, spaCy, Gensim, or similar libraries. 
In general, we found that, qualitatively, LDA produced less meaningful results than NMF, as our topic description file primarily contained filler words, stopwords, and numbers, making it challenging to summarize coherent topics. 
Attempts to refine LDA preprocessing through stopword removal and lemmatization did not sufficiently filter out non-informative words, resulting in reduced topic interpretability compared to the more structured outputs from NMF; with these changes, our visual clustering results did not change significantly. 

We show our final visualization of temporal topic groupings in Figure~\ref{fig:combined_stacked}. 
NMF tends to produce sparse topic representations, meaning that each topic often consists of a smaller set of highly weighted words. 
LDA's probabilistic nature can sometimes result in topics where many words have a non-zero probability, potentially making them less distinct.
There are some interesting patterns in these visual representations, such as the upper orange topic group in our NMF PCA graphs, which appears around October 2022 in Figure 2 and aligns temporally with the sudden NDCG spike measured by our evaluation system.

We also attempted to plot the data using neighborhood manifold techniques, such as t-SNE and UMAP.
However, both resulted in out-of-memory issues despite multiple runs to tune \texttt{n\_components} and \texttt{n\_neighbors}, likely due to a combination of sample size and probability vectors being too large to fit into memory for the optimization routines.


As a note for future reruns of this experiment, quantitative coherence metrics ($ C_{v}, C_{uci}, C_{npmi}$) provide objective, standardized measures of topic quality that eliminate the subjectivity inherent in visual clustering assessments, enabling reproducible model comparison and systematic hyperparameter optimization. 
We mention this because temporal grouping similarities are visually less dynamic in LDA graphs compared to NMF graphs. 
This separation difference is likely due to LDA's inability to thoroughly clean stop words and special characters, resulting in more similar 100-word summaries for each topic. 
Interestingly, despite the lack of token preprocessing in NMF, the resulting plots demonstrate more dynamic shifts in grouping over time. 
NMF may be capturing the underlying structure of our dataset more effectively than LDA. 
Additionally, having more time to experiment with larger samples of the data and varying the number of topics could lead to more distinct grouping outcomes and topic summaries. 
\section{Experimental Methodology}

\subsection{Evaluation System}

\begin{figure}[h]
    \centering
    \includegraphics[width=0.9\textwidth, clip]{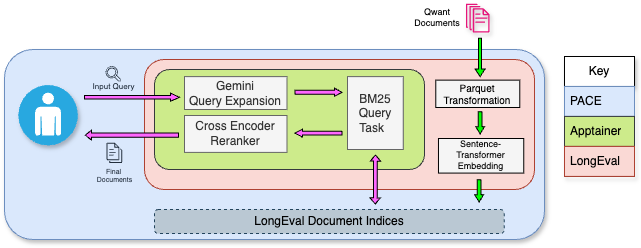}
    \caption{The overall architecture for the two-stage evaluation system.}
    \label{fig:LongEvalArch}
\end{figure}

Our information retrieval engine for document ranking is a preliminary attempt to handle large-scale data processing within high-performance computing environments.
We use a standard two-stage retrieval pipeline to find relevant documents for each query.
Queries are expanded to include a larger number of keywords for a BM25 keyword search to get the 100 most relevant documents.
These documents are then reranked using a cross-encoder before being sent off for evaluation.

\begin{figure}[tbhp]
    \centering
    \begin{subfigure}{\columnwidth}
        \begin{lstlisting}[numbers=left,frame=single]
{{ query_text }}

For each query above, generate a query expansion in French that includes additional relevant terms or phrases.
The query expansion should be no longer than 100 words.
The query engine relies on BM25 and vector search techniques in French.
The output should be a JSON array of objects, each containing the original 'qid' and the expanded 'query'.
\end{lstlisting}
        \caption{Prompt for query expansion.}
        \label{subfig:expansion_prompt}
    \end{subfigure}
    
    \vspace{1em} 

    \begin{subfigure}{\columnwidth}
        \begin{lstlisting}[numbers=left,frame=single]
{
    "type": "array",
    "items": {
        "type": "object",
        "properties": {
            "qid": {
                "type": "string",
                "description": "The query identifier.",
            },
            "query": {
                "type": "string",
                "description": "The text of the query.",
            },
        },
        "required": ["qid", "query"],
        "additionalProperties": False,
    },
}
\end{lstlisting}
        \caption{JSON schema for the structured output.}
        \label{subfig:expansion_schema}
    \end{subfigure}
    \caption{
    The prompt and JSON schema used for query expansion. (a) The prompt instructs a Large Language Model to generate French query expansions suitable for a hybrid search system using BM25 and vector search. (b) The required output schema enforces a structured response containing the query ID and the expanded query text. This allows for reliable integration back into the data processing pipeline.
    }
    \label{fig:query_expansion_setup}
\end{figure}

As illustrated in Figure~\ref{fig:LongEvalArch}, the system begins by ingesting Qwant documents, transforming them into Parquet format, and applying optional sentence-transformer embeddings. 
These preprocessing steps are executed on Georgia Tech's PACE supercomputing cluster, orchestrated via SLURM workload managers~\cite{slurm}, with the processed data stored in a shared directory for downstream retrieval.  
For the retrieval workflow, we employ Pyserini~\cite{pyserini} for keyword searches, with supporting batch processing pipelines implemented using PySpark and Luigi~\cite{luigi}. 
At query time, the system references a precomputed mapping index of Qwant queries to their Gemini-expanded variants.
We use the prompt in Figure~\ref{fig:query_expansion_setup} and ensure that the output adheres to a concrete schema.
All hyperparameters such as temperature are set to the default values as per the Gemini API SDK.

\subsection{Evaluation Metrics}

We evaluate the system against the relevant information retrieval measures for the task. 
The normalized Discounted Cumulative Gain (nDCG) metric is defined as follows:

\begin{equation}
    \begin{array}{ccc}
 nDCG_k = \frac{DCG_k}{IDCG_k} & \quad DCG_k = \sum_{i=1}^{k} \frac{rel_i}{\log_2(i+1)} & \quad IDCG_k = \sum_{i=1}^{k} \frac{rel_i}{\log_2(i+1)}
    \end{array}
\end{equation}

IDCG represents the maximum achievable DCG with the same set of relevance scores but in the perfect ranking order. 
This equation rewards relevant documents appearing early in the ranked list and is especially important in web search contexts. 
We also measure the relative nDCG drop, the formula for which is as follows under the context of lag data:

\begin{equation}
RND = \frac{nDCG_{\text{lag6}} - nDCG_{\text{lag8}}}{nDCG_{\text{lag6}}}
\end{equation}

\section{Results}

\subsection{Retrieval System}

We report the training and test NDCG@10 for our four experiments.
The BM25 experiment yields the top 100 results per query.
Query expansion replaces the original query with terms generated from an LLM.
Reranking utilizes a French-specific reranking sentence transformer (\texttt{antoinelouis/crossencoder-camembert-base-mmarcoFR}) to reweight result sets.
We do not perform fine-tuning on the reranking model.

In the average over NDCG@10 in Table~\ref{tab:experiment_results}, we find that reranking provides the most gain in performance in our pipeline.
When considering ablation, removing the reranking stage reduces performance by 0.11, whereas removing query expanded results increases the score by 0.01. 
The performance pattern holds generally over time.
The reranked results score higher than non-reranked results, while the original queries perform better than expanded queries.

In the scores over time in Table~\ref{tab:monthly_by_experiment}, scores correlate well during the roughly one-year period preceding the last date in the test set.
Before this time, the performance across all models decreases significantly.
We observe that the relative ranking between models follows the average score.

\begin{table}[htbp]
\centering
\caption{Aggregated nDCG@10 Results by Experiment}
\label{tab:experiment_results}
\begin{tabular}{l l l}
\toprule
Experiment & {nDCG@10 Mean} & {nDCG@10 Std Dev} \\
\midrule
bm25-reranked & 0.296 & 0.371 \\
bm25-expanded-reranked & 0.295 & 0.375 \\
bm25 & 0.242 & 0.337 \\
bm25-expanded & 0.194 & 0.314 \\
\bottomrule
\end{tabular}
\end{table}


\begin{table}[htbp]
\centering
\caption{
    Monthly NDCG@10 by Experiment.
    The NULL value in 2022-08 is due to an intermittent processing failure during execution of the pipeline. 
}
\label{tab:monthly_by_experiment}
\begin{tabular}{l l l l l}
\toprule
Date & {BM25} & {BM25-Expanded} & {BM25-Expanded-Reranked} & {BM25-Reranked} \\
\midrule
2022-06 & 0.127 & 0.11 & 0.157 & 0.161 \\
2022-07 & 0.134 & 0.116 & 0.163 & 0.168 \\
2022-08 & 0.141 & 0.123 & {NULL} & 0.176 \\
2022-09 & 0.21 & 0.184 & 0.249 & 0.259 \\
2022-10 & 0.296 & 0.236 & 0.344 & 0.361 \\
2022-11 & 0.292 & 0.226 & 0.34 & 0.367 \\
2022-12 & 0.303 & 0.238 & 0.35 & 0.372 \\
2023-01 & 0.312 & 0.237 & 0.352 & 0.379 \\
2023-02 & 0.31 & 0.251 & 0.356 & 0.378 \\
2023-03 & 0.316 & 0.246 & 0.354 & 0.39 \\
2023-04 & 0.323 & 0.253 & 0.361 & 0.39 \\
2023-05 & 0.327 & 0.255 & 0.368 & 0.395 \\
2023-06 & 0.318 & 0.253 & 0.353 & 0.373 \\
2023-07 & 0.319 & 0.252 & 0.362 & 0.391 \\
2023-08 & 0.284 & 0.223 & 0.316 & 0.34 \\
\bottomrule
\end{tabular}
\end{table}

\begin{figure}[htbp]
    \centering
    \includegraphics[width=0.75\linewidth]{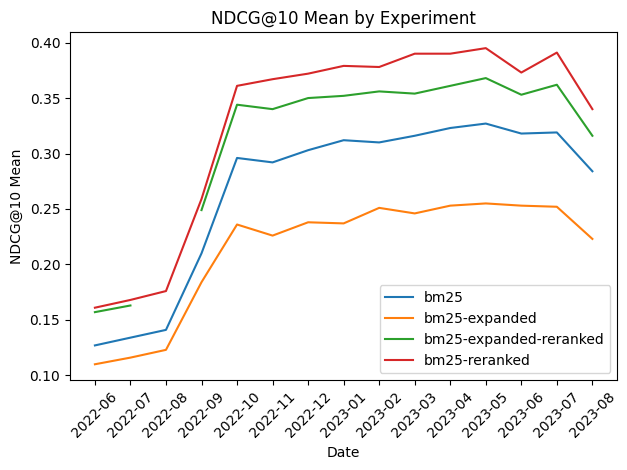}
    \caption{
        NDCG performance over time by experimentation.
        Note that there is a recency bias with a regime change around 2022-09.
    }
    \label{fig:enter-label}
\end{figure}

\section{Discussion}

The three main conclusions we can draw from the results of the retrieval system are that rerankers provide proper relevancy signals over keyword result sets, that our particular query expansion method reduces the performance of searches in this dataset, and that anomalous behavior exists in the system's performance prior to October 2022.

Rerankers have proven effective in practice for IR systems, so it is reassuring to see that the raw BM25 result sets exhibit a performance boost.
Although we did not perform any further tuning on our dataset, we observed a consistent performance gap between systems that remains constant over time.

Despite query expansion underperforming relative to the original queries, the reranking mechanism can account for those differences.
What this implies, then, is that, on average, the query results in a similar number of relevant documents in the top 100. However, in the expanded queries, important frequency-adjusted keywords are lost due to factors such as repetition.
The reranking semantic space can overcome the limitations of keyword-based retrieval, thus bringing more relevant documents to the top.
We need to recompute the MAP scores to strengthen this conjecture.
However, this would help explain the performance gap we are seeing with the query expanded results.

We also note that during one of the batches of query expansion prompting, we encountered content-based restrictions due to the appearance of explicit content search terms.
These rows appear in the last 1/100th of queries sorted by their query IDs in numerical order.
We use DeepSeek R1 to generate the last batch of query expansions.
We found that OpenAI models result in a similar set of problems around content filtering, making it challenging to perform query expansion when using an external API on a model that is not self-hosted.
In the future, we might instead rely on self-hosted model like Llama or Gemma for query expansion.

Lastly, we make note of the regime change one year prior to the last element in the test set.
While there are minor fluctuations that warrant further study regarding the performance gap between models, this shift in performance is challenging to explain.
One possibility is that the distribution of tokens changes significantly in the first few months of the dataset — for example if the document set had a larger distribution of non-French documents, which could cause issues with the French analyzers used by Lucene.
Another possibility is that many of the queries have temporal saliency to a particular event in time, i.e., a non-trivial number of queries that reference an event that occurs in September or October 2022.

\section{Future Work}

Future work would involve incorporating dense retrieval into the pipeline.
Embedding the entire set of documents is a computationally expensive endeavor, but it would likely significantly improve retrieval results.
Models like Nomic Embed v2 demonstrate state-of-the-art retrieval performance across multiple languages and domains and would likely prove viable as a lens into the challenges posed by LongEval.
We provide some analysis into the running time for the entire dataset in Section~\ref{sec:dataset-stats}, and reiterate that each date split takes about a day of Nvidia v100 GPU-time.

We would also like to dive deeper into the topic modeling exploratory analysis.
The generative keyword-based approach, summarized by a stronger LLM, provides a cost-effective way to organize documents into non-overlapping clusters and offers deeper insights than geometric methods like K-means.
While we have examined changes in distribution over time, we would like to see how retrieval scores change over time for popular topics identified at particular snapshots in time.

Future direction might also include retrieval methods that diverge from the typical solutions that involve keyword and dense-embedding searches.
Given the network structure of the web, it seems natural to perform retrieval based on implicit link structures between pages.
A two-stage retrieval pipeline like ours can be augmented with node centrality measures, such as PageRank, to help reweight the relevancy of important documents based on the link structure. 
Methods like K-core decomposition can also help prune documents that are likely to be high-frequency noise in the graph.
Graph theoretic analysis techniques also apply to the semantic k-NN graph of a dense embedding model. Although the link structure is different, the mechanics of the algorithms on the pipeline would be interesting to explore in the context of the temporal evolution of systems.

\section{Conclusions}

Our experiments reveal the effectiveness of pre-trained reranking methods in enhancing retrieval performance within keyword-based search systems.
We uncover a temporal anomaly in the search system, where query performance degrades in partitions older than a year. 
LDA and NMF both offer insight into data evolution, with NMF yielding more apparent cluster separation for our temporal-aware IR system.
Existing techniques for retrieval and analysis still present many opportunities for refinement in future LongEval labs, which would be a critical medium for advancing resilient, time-aware information retrieval.

\section*{Acknowledgements}

Thank you to the DS@GT CLEF team for their support. This research was also supported in part through cyberinfrastructure resources and services provided by the Partnership for an Advanced Computing Environment (PACE) at the Georgia Institute of Technology, Atlanta, Georgia, USA.

\section*{Declaration on Generative AI}

During the preparation of this work, the author(s) used Gemini Pro and Grammarly in order to: Abstract drafting, formatting assistance, grammar and spelling check. After using these tool(s)/service(s), the author(s) reviewed and edited the content as needed and take(s) full responsibility for the publication’s content. 

\bibliography{main}


\end{document}